\documentclass[useAMS,usenatbib,twocolumn]{mnras}
\usepackage[figuresright]{rotating}
\usepackage{lscape}

\usepackage{graphicx}
\usepackage{epstopdf}
\usepackage{amsmath}
\usepackage{amssymb}
\usepackage{multirow}
\usepackage{setspace}
\usepackage{bigdelim}
\usepackage{bigstrut}
\usepackage[rflt]{floatflt}
\usepackage{wrapfig}
\usepackage{caption}
\usepackage{lipsum}

\usepackage{multicol}

\usepackage{indentfirst}

\title[Constraining $f_\mathrm{esc}$ with SKA]{The Escape Fraction of Ionizing Photons During the Epoch of Reionization: observability with the Square Kilometre Array}
\author[Seiler et al]{Jacob Seiler$^{1,2}$\thanks{E-mail:
		jseiler@swin.edu.au}, Anne Hutter$^{1,2,3}$, Manodeep Sinha$^{1,2}$, Darren Croton$^{1,2}$\\
\\
$^{1}$Centre for Astrophysics \& Supercomputing, Swinburne University of Technology, Hawthorn, Victoria 3122, Australia\\
$^{2}$ARC Centre of Excellence for All Sky Astrophysics in 3 Dimensions (ASTRO 3D)\\
$^{3}$Kapteyn Astronomical Institute, University of Groningen, PO Box 800, 9700 AV Groningen, The Netherlands}

\begin{document}

\pagerange{\pageref{firstpage}--\pageref{lastpage}} \pubyear{2002}

\maketitle

\label{firstpage}

\begin{abstract}

One of the most important parameters in characterizing the Epoch of Reionization, the escape fraction of ionizing photons, $f_\mathrm{esc}$, remains unconstrained both observationally and theoretically.  With recent work highlighting the impact of galaxy-scale feedback on the instantaneous value of $f_\mathrm{esc}$, it is important to develop a model in which reionization is self-consistently coupled to galaxy evolution. In this work, we present such a model and explore how physically motivated functional forms of $f_\mathrm{esc}$ affect the evolution of ionized hydrogen within the intergalactic medium.  Using the $21$cm power spectrum evolution, we investigate the likelihood of observationally distinguishing between a constant $f_\mathrm{esc}$ and other models that depend upon different forms of galaxy feedback. We find that changing the underlying connection between $f_\mathrm{esc}$ and galaxy feedback drastically alters the large-scale $21$cm power. The upcoming Square Kilometre Array Low Frequency instrument possesses the sensitivity to differentiate between our models at a fixed optical depth, requiring only $200$ hours of integration time focused on redshifts $z = 7.5-8.5$.  Generalizing these results to account for a varying optical depth will require multiple $800$ hour observations spanning redshifts $z = 7-10$. This presents an exciting opportunity to observationally constrain one of the most elusive parameters during the Epoch of Reionization.

\end{abstract}


\section{Introduction}

The Epoch of Reionization, which is completed by redshift $z \sim 6$ \citep{Fan2006,Becker2015}, represents the final phase transition of the Universe from a neutral, post-recombination state to the highly ionized one that we observe today.  As the first stars form they radiate photons which gradually ionize the neutral hydrogen within the intergalactic medium (IGM).   During reionization an intense ultraviolet background (UVB) builds up, photoheating the IGM and evaporating baryons within low-mass dark matter halos \citep{Shapiro2004}.  For the galaxies residing in these halos, star formation is severely suppressed and often halted.  As these galaxies provide the starting conditions for the Universe we see today, it is paramount to understand the interplay between the Epoch of Reionization and galaxy evolution.

An important parameter for understanding the Epoch of Reionization is the fraction of hydrogen ionizing photons that escape from galaxies into the IGM, $f_\mathrm{esc}$. This parameter strongly dictates the speed and duration of reionization, in addition to affecting the size and topology of the ionized regions.  Since the ionizing flux is absorbed by the intervening neutral IGM, observationally measuring $f_\mathrm{esc}$ during the Epoch of Reionization is not possible.  At lower redshifts where direct measurement is not precluded, there is some consensus on the value of $f_\mathrm{esc}$. Within this regime (redshift $0 < z < 1.5$), escape fractions of the order of $\sim 0.01$-$0.05$ have been predominantly observed \citep[e.g.,][]{Cowie2009,Grimes2009,Siana2010}. As the redshift increases to $z \sim 4$, $f_\mathrm{esc}$ has been observed to range from $0.10$ to $0.20$ \citep[e.g.,][]{Vanzella2010,Guaita2016,Grazian2016, Steidel2018}.  However, there also exist a number of measurements that challenge these trends.  For example, \cite{Vanzella2016}, \cite{Bian2017} and \cite{Vanzella2018} observe galaxies with $f_\mathrm{esc}$ lower limits of $0.50$, $0.28$ and $0.60$ at redshifts $z = 3.2$, $z = 2.5$, and $z = 4$, respectively.  Such outlying observations highlight the uncertainty surrounding exactly how $f_\mathrm{esc}$ varies with galaxy properties.

Based mostly on radiation-hydrodynamical simulations, there exists a growing body of work that highlights the importance of galaxy-scale processes in regulating the instantaneous value of $f_\mathrm{esc}$.  \cite{Paardekooper2015} finds that the spatial distribution of gas inside a halo can dictate the values of $f_\mathrm{esc}$.  Importantly, they find that processes such as supernovae feedback are able to disperse dense gas clouds and permit the easy escape of ionizing photons.  This was confirmed by \cite{Kimm2017} and \cite{Trebitsch2017}, where the simulations show that this process is most efficient in low-mass galaxies where gas is easily expelled due to the low gravitational potential.  This picture, that $f_\mathrm{esc}$ should be largest for lowest mass galaxies, is consistent with previous works highlighting that $f_\mathrm{esc}$ scales negatively with dark matter halo mass \citep[e.g.,][]{Ferrara2013,Kimm2014,Wise2014}.  However, such a conclusion is not so clear-cut, with a handful of models finding $f_\mathrm{esc}$ to instead scale positively with halo mass \citep[e.g.,][]{Gnedin2008,Wise2009}.  Taking both the observations and simulations together depicts $f_\mathrm{esc}$ as a highly unconstrained parameter that depends sensitively on the properties of the host dark matter halo and the underlying feedback processes within the galaxy itself.

In an effort to quantify the impact of $f_\mathrm{esc}$ on parameters such as the duration of reionization and topology of ionized regions, a number of works have analysed reionization models under different values and functional forms of $f_\mathrm{esc}$.  \cite{Bauer2015} post-process a radiative transfer scheme with the hydrodynamic Illustris simulation and show that the duration of reionization varies between $190$ and $340$ Myr if $f_\mathrm{esc}$ scales with redshift. Similar results are echoed by \cite{Doussot2019} who use a self-consistent radiative-hydrodynamic simulation to follow the evolution of ionized hydrogen for an $f_\mathrm{esc}$ value that scales linearly or quadratically with redshift.  There have been few works investigating the impact of $f_\mathrm{esc}$ on the topology of ionized regions.  The most relevant is \cite{Kim2013} who show that by employing $f_\mathrm{esc}$ values that vary with halo mass or redshift, the resulting size and distribution of ionized regions differ noticeably.   Such a result highlights that the functional form of $f_\mathrm{esc}$ plays a key role in setting the topology of ionized hydrogen during reionization.

One of the most promising avenues in detecting the Epoch of Reionization and its topology is through measurement of the low-frequency $21$cm hydrogen line \citep[see reviews by e.g.,][]{Furlanetto2006,Pritchard2012} with radio telescopes such as the Murchison Widefield Array \citep[MWA;][]{Tingay2013}, Square Kilometre Array \citep[SKA;][]{Carilli2004}, Low Frequency Array \citep[LOFAR;][]{vanHaarlem2013}, and Hydrogen Epoch of Reionization Array \citep[HERA;][]{DeBoer2017}.  Importantly, these instruments will measure the $21$cm transition on a range of scales with the signal intensity depending upon the presence of neutral hydrogen.  Hence, to fully exploit the scientific power of these next-generation telescopes, we require accurate models detailing the evolution of ionized hydrogen during the Epoch of Reionization. 

The coupling between $f_\mathrm{esc}$ and galaxy feedback should significantly impact both the duration of reionization and the topology of ionized regions.  Furthermore, this coupling could have important consequences for the $21$cm power spectrum that could affect the detectability of the Epoch of Reionization. In this paper we attempt to self-consistently model such coupling with the Reionization using Semi-Analytic Galaxy Evolution (\texttt{RSAGE}) model\footnote{Available at \url{https://github.com/jacobseiler/rsage}}.  Similar to work by the Dark Ages, Reionization And Galaxy-formation Observables Numerical Simulation (DRAGONS) team\footnote{See \url{http://dragons.ph.unimelb.edu.au/} and \cite{Mutch2016}.}, we use galaxies as the source of ionizing photons and follow the evolution of ionized hydrogen, self-consistently accounting for reionization feedback by suppressing the infall of baryons onto low-mass galaxies.  Compared to the DRAGONS model \texttt{Meraxes} \citep[described by][]{Mutch2016}, \texttt{RSAGE} uses synthetic spectra to track the ionizing emissivity of each star formation event through time, rather than relying solely on the stellar mass history and imposing a fixed number of photons per stellar baryon.  This method better captures the evolution of O- and B-type stars which contribute most of the hydrogen ionizing photons.  To simulate the evolution of ionized hydrogen during the Epoch of Reionization, \texttt{RSAGE} is uses the newly developed semi-numerical model \texttt{cifog} \citep{Hutter2018}\footnote{Available at \url{https://github.com/annehutter/grid-model}}.  Unlike other semi-numerical models such as \texttt{21cmFAST} \citep{Mesinger2011} and \texttt{Simfast21} \citep{Santos2010,Hassan2016}, \texttt{cifog} does not evolve a density field using the Zel'dovich approximation \citep{Zeldovich1970} to determine the fraction of collapsed matter and resulting ionizing emissivity.  Instead, \texttt{cifog} uses an input list of ionizing sources thereby aligning naturally with the \texttt{RSAGE} framework which focuses on galaxies as the sources of ionizing photons.

Using \texttt{RSAGE}, we calculate $f_\mathrm{esc}$ values that depend uniquely upon galaxy properties, quantifying the impact $f_\mathrm{esc}$ has on the optical depth, duration of reionization, and $21$cm power spectrum.  Importantly, the efficiency with which the \texttt{RSAGE} model is able to simulate both galaxy evolution and reionization offers us the ability to investigate a variety of different functional forms of $f_\mathrm{esc}$, an advantage not permitted by the computationally expensive radiation-hydrodynamic works such as the Cosmic Reionization on Computers \citep{Gnedin2014} and Cosmic Dawn \citep{Ocvirk2016} simulations.  We propose a new diagnostic plot that tracks the large- and small-scale $21$cm power throughout reionization.  This plot has the potential to distinguish between different functional forms of $f_\mathrm{esc}$ and identifies the SKA observational sweet spot where such differences are maximal.  

The paper is organized as follows. In Section \ref{sec:Modelling} we outline the \texttt{RSAGE} model, highlighting the semi-numerical scheme we use to follow the evolution of ionized hydrogen within the IGM. Section \ref{sec:Models} provides an overview of the $f_\mathrm{esc}$ models we analyse in this work.  In Section \ref{sec:History} we compare the history and duration of reionization for each $f_\mathrm{esc}$ model. We then explore avenues to distinguish between $f_\mathrm{esc}$ models in Section \ref{sec:Morphology}, where we show the differences in ionized region topology and evolution of the $21$cm power spectra. We provide an overview of the model caveats in Section \ref{sec:Discussion} and conclude in Section \ref{sec:Conclusion}. Throughout this paper we adopt the cosmological values $\left(h, \Omega_\mathrm{m}, \Omega_\Lambda, \sigma_8, n_s\right)= \left(0.681, 0.302, 0.698, 0.828, 0.96\right)$ consistent with \cite{Planck2016b}, and use a \cite{Chabrier2003} initial mass function (IMF) where required.

\section{Simulating Reionization}  \label{sec:Modelling}

In this section, we summarize our simulation and the modelling procedures.  We begin with a description of the collisionless N-body simulation used as an input, then provide an overview of our semi-analytic galaxy formation model adapted to high redshift.  In particular, we elaborate on the self-consistent coupling between galaxy evolution and ionized hydrogen during the Epoch of Reionization

\subsection{N-Body Simulation}

In order to capture the evolution of galaxies over cosmic time, we derive the growth and merger histories of their host dark matter halos from the N-Body simulation \textit{Kali} \citep{Seiler2018}.  \textit{Kali} contains $2400^3$ dark matter particles within a $160$Mpc side box, resolving halos of mass $\sim$$4 \times 10^8\mathrm{M}_\odot$ with $32$ particles.   The particles were evolved using \texttt{GADGET-3} \citep{Springel2005} with $98$ snapshots of data stored between redshifts $z = 30$ and $z=5.5$ in $10$Myr spaced intervals.  We refer the interested reader to \cite{Seiler2018} for more information regarding the \textit{Kali} initial conditions and merger tree construction.

\subsection{Semi-Analytic Galaxy Modeling} \label{sec:SAM}

\begin{figure*}
	\includegraphics[scale = 0.4]{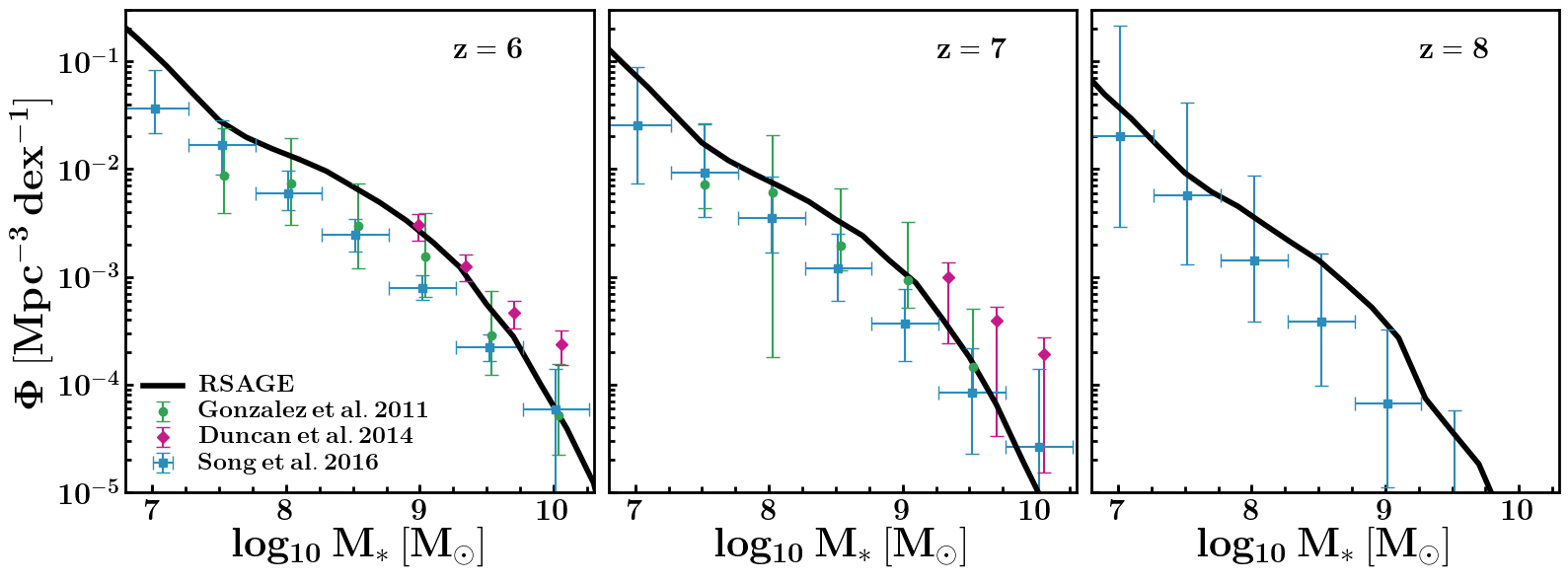}
	\caption{Stellar mass function at redshift $z=6, 7, 8$ using the \texttt{RSAGE} model.  This model differs from C16 by using delayed supernova feedback and includes self-consistent reionization feedback.  For clarity, we show here only the Constant model as all other models (Section \ref{sec:Models}) have very similar values by choice.  Parameters were chosen to match observations from \citet{Gonzalez2011}, \citet{Duncan2014} and \citet{Song2016}.  All observations have been corrected to a Chabrier IMF and a Hubble parameter $h = 0.698$.}
	\label{fig:StellarMass}
\end{figure*}

Within \texttt{RSAGE}, the evolution of galaxies over cosmic time uses the Semi-Analytic Galaxy Evolution (\texttt{SAGE}) model of \citet[][hereafter C16]{Croton2016} as a base.  The \texttt{SAGE} model includes baryonic accretion, cooling, star formation, gas ejection due to supernova feedback, active galactic nuclei feedback through `radio mode' heating and `quasar mode' gas ejection, and galaxy mergers.  In this work, the only galaxy model prescriptions we have changed with respect to C16 are the supernova and reionization feedback schemes, explained below.

In the C16 model, supernova feedback is applied instantaneously.  That is, following a star formation episode, a fraction of stars immediately explode, reheating cold gas and ejecting hot gas.  Whilst such an approximation is valid at low redshift where the time between snapshots is larger than the lifetime of a supernova candidate star, the same is not true during the Epoch of Reionization.  Instead, we closely follow \cite{Mutch2016} and release energy from supernova activity gradually over a number of subsequent snapshots. This results in a much smoother and physically motivated ejection history for each galaxy.
   
To model the effect of reionization on the evolution of galaxies, C16 implement an analytic prescription using fits to the hydrodynamic simulations of \cite{Gnedin2000}.  Importantly, this prescription adopts the parametrization of \cite{Kravstov2004} which uses the redshift at which the first HII regions overlap and the redshift when reionization is completed.  As this parametrization is universal, it ignores the effect of inhomogeneous reionization and switches reionization `on' for all galaxies regardless of mass or environment.  We describe our new reionization feedback scheme in Section \ref{sec:SelfCon}.

Whilst C16 utilized a primary and secondary set of constraints to choose their fiducial set of parameters, here we adjust the galaxy evolution parameters manually to match the high-redshift stellar mass function using \cite{Gonzalez2011}, \cite{Duncan2014} and \cite{Song2016} between $z=6$-$8$ (Figure \ref{fig:StellarMass}). This involved altering the following C16 parameters: the star formation rate efficiency $\alpha_\mathrm{SF}$ from $0.05$ to $0.03$ and the quasar mode ejection efficiency $\kappa_\mathrm{Q}$ from $0.005$ to $0.02$.  We use the \cite{Mutch2016} mass loading and energy coupling constants for supernova feedback and note that these values are scaled depending upon the host halo properties.  Even with such minimal changes to the parameter values, Figure \ref{fig:StellarMass} shows that the stellar mass function matches the observations well over all redshifts, highlighting \texttt{RSAGE}'s robustness in modelling galaxy evolution during the Epoch of Reionization.

\subsection{Self-Consistent Reionization} \label{sec:SelfCon}

The initial works of \cite{Couchman1986} and \cite{Efstathiou1992} highlighted that the presence of an ultraviolet background (UVB) has significant consequences for galaxy evolution.  By photoheating gas within the IGM to temperatures above $10^4$K, the UVB acts to increase the Jeans mass for galaxies located within ionized regions, causing a severe decrease in star formation.  Furthermore, as ionization fronts sweep across the IGM, gas within low-mass ($\sim$ $10^7$-$ 10^8$M$_\odot$) halos is photoheated and subsequently evaporated \citep{Shapiro2004}.  Through these two mechanisms, galaxies embedded within the ionized IGM can have their evolution severely stunted.  As the galaxies formed during this early epoch provide the initial conditions for subsequent galaxy assembly, understanding the evolution of ionized gas during the Epoch of Reionization and its impact on galaxy evolution is critical.

To address this, \texttt{RSAGE} includes a coupled treatment of reionization and its associated feedback on galaxy evolution.  In our model, we use the semi-numerical code \texttt{cifog} \citep{Hutter2018} to generate an inhomogeneous UVB and follow the evolution of ionized hydrogen during the Epoch of Reionization.  By using the galaxies simulated from \texttt{RSAGE} as ionizing sources, we are able to follow both galaxy formation and the progression of reionization in a self-consistent manner.
Motivated by work such as \cite{Iliev2007}, \cite{Iliev2012} and further extensions by \cite{Mutch2016} for the \texttt{Meraxes} model, we follow reionization self-consistently by iterating through the \textit{Kali} simulation snapshots and implementing the following algorithm:

\begin{enumerate}
    \item{Galaxies are evolved to the end of the current snapshot using the \texttt{RSAGE} galaxy evolution model as described in Section \ref{sec:SAM}. Using each star formation history, the number of ionizing photons produced by each galaxy is calculated.  Combined with the escape fraction of ionizing photons (Section \ref{sec:Models}), \texttt{RSAGE} then generates a grid of ionizing sources.}
	\item{By comparing the number of HI ionizing photons with the number of neutral hydrogen atoms and adjusting for recombinations and self-shielding, \texttt{cifog} determines the ionization state and local UVB strength within each grid cell.}
	\item{\texttt{RSAGE} tracks the redshift at which each grid cell is ionized and generates a suppression modifier for dark matter halos within these cells. The baryonic content of halos within ionized regions is suppressed using this modifier and \texttt{RSAGE} proceeds to the next snapshot by cycling back to step (i).}
\end{enumerate}

In the following sub-sections we elaborate on each of these three steps. 

\subsubsection{Ionizing Photons}

The number of ionizing photons that escape from each galaxy in the simulation is determined by the number of ionizing photons intrinsically produced ($N_{\gamma,i}$) and the escape fraction $f_\mathrm{esc,i}$ (see Section \ref{sec:Models} for our models of $f_\mathrm{esc}$),

\begin{equation}
N_\mathrm{ion} = \sum_i^{N_s} f_{\mathrm{esc},i} N_{\gamma,i},
\end{equation}

\noindent where $N_s$ is the number of ionizing sources (i.e., the number of galaxies).  

Previously, we linked the value of $N_\gamma$ to a galaxy's star formation rate across each snapshot \citep{Seiler2018}.  However, as star formation can be completely shut down for a number of snapshots due to supernova or quasar feedback, galaxies with non-zero stellar mass were marked as producing no ionizing photons.  In our updated version of \texttt{RSAGE}, we instead link the number of ionizing photons produced by each galaxy to its star formation \textit{history}.  In a similar manner to the delayed supernova scheme, we store the past $100$Myr of star formation in step sizes of $1$Myr\footnote{The \texttt{SAGE} model supports the use of sub-steps.  Hence, whilst the time between \textit{Kali} snapshots is $10$Myr, galaxies are evolved in $1$Myr time steps}.  By using spectra generated from \texttt{STARBURST99} \citep{Leitherer1999} with a \cite{Chabrier2003} initial mass function, we determine the number of ionizing photons, $N_\gamma \left(M_\mathrm{burst}\right)$, emitted from an instantaneous starburst that formed mass $M_\mathrm{burst}$ of stars.  In practice, as the number of ionizing photons produced in the burst scales linearly with $\log{M_\mathrm{burst}}$, we run \texttt{STARBURST99} for $M_\mathrm{burst} = 10^6 \mathrm{M}_\odot$ and scale our results for any arbitrary star formation episode.  The number of ionizing photons emitted from a galaxy at any given time $t$ is then given by

\begin{align}
N_\gamma\left(t\right) & = \sum_{t_i = 1, 2, 3, ...}^{100} N_\gamma\left(M_\mathrm{burst} \left(t_i\right)\right), \tag*{}\\
& = \sum_{t_i = 1, 2, 3, ...}^{100} \frac{M_\mathrm{burst} \left(t_i\right)}{10^6 \mathrm{M}_\odot}, \\ & \quad \quad \quad \quad \times N_\gamma\left(M_\mathrm{burst} = 10^6 \mathrm{M}_\odot \left(t_i\right)\right), \tag*{}
\end{align}   

\noindent where $M_\mathrm{burst}\left(t_i\right)$ is the mass of stars formed $t_i$Myr ago.

\subsubsection{The Reionization of Neutral Hydrogen} \label{sec:cifog}

Once the number of ionizing photons for each galaxy has been determined, we follow the evolution of ionized gas using the grid-based code \texttt{cifog}, a newly developed, publicly available\footnote{\url{https://github.com/annehutter/grid-model}} MPI-parallelised semi-numerical code which models the ionization of both hydrogen and helium. We summarize here the main features of \texttt{cifog} and refer the interested reader to \cite{Hutter2018} for a detailed description of the code.

In order to flag ionized regions within the IGM, \texttt{cifog} uses the excursion set formalism approach of \cite{Furlanetto2004} in which a region is flagged as ionized ($\chi_\mathrm{HI} = 0$) if the number of ionizing photons exceeds the number of absorptions; otherwise it is marked as neutral ($\chi_\mathrm{HI} = 1$).  Beginning at large radii and progressing towards small scales, the criterion on whether the central cell\footnote{Some semi-numerical models, such as \texttt{Simfast21} \citep{Santos2010}, mark the entire spherical region as ionized if equation \ref{eq:Reionization} is satisfied.  Here we only mark the central cell as ionized.  We refer interested readers to \cite{Hutter2018} for discussion regarding the impact of difference flagging schemes.} is flagged as ionized at redshift $z$ is,

\begin{equation}
\begin{split}
\int_{z}^{\infty}N_\mathrm{ion}\left(z\right) \: dz &  \geq \int_{z}^{\infty}  N_\mathrm{abs}\left(z\right) \: dz, \\
& \geq \langle n_\mathrm{H,0}\rangle_R V_\mathrm{cell}\left[1 + \int_{z}^{\infty} \langle N_\mathrm{rec}\rangle_R\left(z\right) \: dz\right], 
\end{split}
\label{eq:Reionization}
\end{equation}

\noindent where $n_\mathrm{H,0}$ is the hydrogen number density today,  $V_\mathrm{cell}$ is the comoving volume of a grid cell, $N_\mathrm{rec}$ is the number of recombinations and $\langle \rangle_R$ denotes the average over the spherical region with radius $R$.  Provided the radius over which the ionizing photons and recombinations are counted is large enough, this method automatically accounts for ionization from neighbouring or distant bright sources.

To calculate the suppression of baryonic gas infall (Section \ref{sec:Suppression}), \texttt{cifog} must calculate the spatially dependent photoionization rate, $\Gamma_\mathrm{HI} \left(\mathbf{x},z\right)$.  The photoionization rate represents the number of ionization events per unit time and is a function of the ionizing flux incident upon each grid cell,

\begin{equation}
\Gamma_\mathrm{HI}\left(\mathbf{x},z\right) \propto \sum_{i = 0}^{N_s}\frac{N_{\gamma,i}}{4\pi\left|\mathbf{x}-\mathbf{x}_\mathrm{i}\right|^2} e^{-\frac{\left|\mathbf{x}-\mathbf{x}_\mathrm{i}\right|}{\lambda_\mathrm{mfp}}},
\label{eq:Photoionization}
\end{equation}

\noindent where $\mathbf{x} = \left(x,y,z\right)$ is the position of the ionizing source and we perform the sum over all $N_s$ sources. $\lambda_\mathrm{mfp}$ is the median value of the mean free path of ionizing photons, which, during reionization, is given by the size of ionized regions (i.e., the largest scale, $R$,  at which the excursion set formalism marks a region as being ionized). Towards the final stages of reionization, when ionized regions begin to merge, $\lambda_\mathrm{mfp}$ is instead given by the distance between self-shielded regions.

\subsubsection{Suppression of Baryonic Infall} \label{sec:Suppression}

As mentioned previously, the presence of an UVB can suppress star-formation by increasing the Jeans mass and photoevaporating gas within low-mass halos. By running a suite of 1D cosmological collapse simulations, \cite{Sobacchi2013} capture these effects through their impact on the universal halo baryon fraction, $f_\mathrm{b}$.  They provide a parametrization whereby halos within ionized regions have their baryon fraction suppressed by a factor of $f_\mathrm{mod}$,

\begin{equation}
f_\mathrm{mod} \left(M_\mathrm{H}\right) = 2^{-2M_\mathrm{crit}/M_\mathrm{H}},
\label{eq:fmod}
\end{equation}

\noindent where $M_\mathrm{H}$ is the halo mass and $0 \leq f_\mathrm{mod} \leq 1$.  Here $M_\mathrm{crit}$ is defined as the halo mass that is able to retain half of its baryons in the presence of an UVB (i.e., $f_\mathrm{mod} = 0.5$).  This value depends on the halo mass, UVB intensity ($\Gamma_\mathrm{HI}$), current redshift ($z$) and the redshift at which the surrounding IGM was ionized ($z_\mathrm{reion}$),

\begin{equation}
M_\mathrm{crit} = M_0\Gamma_\mathrm{HI}^a\left(\frac{1 + z}{10}\right)^b\left[1 - \left(\frac{1 + z}{1 + z_\mathrm{reion}}\right)^c\right]^d,
\label{eq:Mcrit}
\end{equation}

\noindent where $\left(M_0, a, b, c, d\right)$ are fitting parameters of the \cite{Sobacchi2013} model and found to be $\left(2.8 \times 10^9 M_\odot, 0.17, -2.1, 2.0, 2.5\right)$.

Similar to the \texttt{Meraxes} model outlined in \cite{Mutch2016}, \texttt{RSAGE} tracks the redshift at which each cell in the simulation box becomes ionized.  From Equations \ref{eq:Photoionization} and \ref{eq:Mcrit}, we generate a list of baryon modifiers (Equation \ref{eq:fmod}) for all halos within ionized regions and suppress the baryonic content for the hosted galaxies according to the photoionization rate in each cell (Equation \ref{eq:Photoionization}).

\section{The Ionizing Escape Fraction}  \label{sec:Models}

\begin{table*} 
	\begin{tabular}{|c|p{35mm}|c|p{25mm}|}
		\hline
		\textbf{Model} & \textbf{Description} & $\mathbf{f_\mathrm{esc}}$ \textbf{Form} & \textbf{Calibration Values} \\ \hline \hline
		Constant & Escape fraction is constant for all galaxies across cosmic time & $f_\mathrm{esc} = \mathrm{Constant}$ & $\mathrm{Constant} = 0.20$ \\ \hline
		MH-Neg & Escape fraction scales inversely as a function of halo mass & $\log_{10} f_\mathrm{esc} = \log_{10}f_\mathrm{esc,low} - \left[ \frac{\log_{10}\frac{M_\mathrm{H}}{M_\mathrm{H,low}}}{\log_{10}\frac{M_\mathrm{H}}{M_\mathrm{H,high}}} \log_{10}\frac{f_\mathrm{esc,low}}{f_\mathrm{esc,high}}\right]$  & $\begin{aligned}[t]M_\mathrm{low} & = 10^5 \mathrm{M}_\odot,\\ M_\mathrm{high} &= 10^{12} \mathrm{M}_\odot,\\ f_\mathrm{esc,low} &= 0.99,\\ f_\mathrm{esc,high} &= 0.10\end{aligned}$ \\ \hline
		MH-Pos & Escape fractional scales proportionally as a function of halo mass & $\log_{10} \left(1 -f_\mathrm{esc}\right) = \log_{10}\left(1 -f_\mathrm{esc,low}\right) - \left[ \frac{\log_{10}\frac{M_\mathrm{H}}{M_\mathrm{H,low}}}{\log_{10}\frac{M_\mathrm{H}}{M_\mathrm{H,high}}} \log_{10}\frac{1 - f_\mathrm{esc,low}}{1 - f_\mathrm{esc,high}}\right]$ & $\begin{aligned}[t]M_\mathrm{low} & = 10^8 \mathrm{M}_\odot,\\ M_\mathrm{high} &= 10^{12} \mathrm{M}_\odot,\\ f_\mathrm{esc,low} &= 0.01,\\ f_\mathrm{esc,high} &= 0.40\end{aligned}$ \\ \hline
		Ejected & Escape fraction scales proportionally to the fraction of galaxy baryons in the ejected reservoir ($f_\mathrm{ej}$) & $f_\mathrm{esc} = \alpha f_\mathrm{ej} + \beta$ & $\begin{aligned}[t]\alpha &= 0.30,\\\beta &= 0.00.\end{aligned}$ \\ \hline
		SFR & Escape fraction scales with the star formation rate of the galaxy & $f_\mathrm{esc} = \frac{\delta}{1 + \exp{\left(-\alpha\left(\log_{10}\mathrm{SFR} - \beta\right)\right)}}$ & $\begin{aligned}[t]\alpha &= 1.00,\\\beta &= 1.50,\\\delta &=1.00.\end{aligned}$ \\ \hline \hline
	\end{tabular}
	\caption{Summary of ionizing escape fraction ($f_\mathrm{esc}$) models.  Each model is calibrated against observations of the stellar mass function from redshift $z=6-8$ \citep{Gonzalez2011,Duncan2014,Song2016}, inferences of the ionizing emissivity from redshift $z=15-6$ \citep{Bouwens2015} and the Thomson optical depth \citep{Planck2018}.  For the MH-Neg and MH-Pos models, $M_\mathrm{low}$, $M_\mathrm{high}$, $f_\mathrm{esc,low}$ and $f_\mathrm{esc,high}$ represent the minimum (maximum) halo mass and the corresponding minimum (maximum) escape fraction for galaxies within these halos.  For halo above (below) these value, $f_\mathrm{esc}$ is set to $f_\mathrm{esc,low}$ ($f_\mathrm{esc,high}$).}
	\label{tab:Models}
\end{table*}

In this section, we describe our different models of the ionizing escape fraction, each depending on a different galaxy property or process. We provide a summary in Table \ref{tab:Models}.  Whilst our coupled model offers the ability to investigate the effect of galaxy formation physics on reionization, here we focus explicitly on how the functional form of $f_\mathrm{esc}$ impacts the timing and topology of reionization.   Hence, we use identical parameters for the galaxy evolution aspect of \texttt{RSAGE} (Section \ref{sec:SAM}) for each model. The free parameters for each model (shown in the `Calibration Values' column of Table \ref{tab:Models}) are adjusted to ensure that the galaxy stellar mass function from redshift $z=6$ to $8$ matches the observations of \cite{Gonzalez2011}, \cite{Duncan2014} and \cite{Song2016}.  We also ensure the ionizing emissivity from redshift $z=6$ to $15$ matches the inferences of \cite{Bouwens2015} and the Thomson optical depth matches the measurements of \cite{Planck2018}.

\vspace*{-0.5mm}

\noindent \textbf{Constant}:  This fiducial model uses a constant value of $f_\mathrm{esc}$ for all galaxies over cosmic time,
\begin{equation}
f_\mathrm{esc} = \mathrm{Constant}.
\end{equation}
\\
\textbf{MH-Neg}:  By calculating the evolution of ionization fronts during the Epoch of Reionization, \cite{Ferrara2013} find that the shallower potential of low-mass halos allows rapid ionization of the interstellar medium due to the decreased number of recombinations.  In turn, this allows subsequent generations of ionizing photons to escape into the IGM more easily. This model follows works such as \cite{Kimm2017} who allow the escape fraction to scale negatively as a power law with dark matter halo mass,

\begin{equation}
\log_{10} f_\mathrm{esc} = \log_{10}f_\mathrm{esc,low} - \left[ \frac{\log_{10}\frac{M_\mathrm{H}}{M_\mathrm{H,low}}}{\log_{10}\frac{M_\mathrm{H}}{M_\mathrm{H,high}}} \log_{10}\frac{f_\mathrm{esc,low}}{f_\mathrm{esc,high}}\right].
\end{equation}

\noindent The fixed points ($M_\mathrm{low}$, $f_\mathrm{esc,low}$) and ($M_\mathrm{high}$, $f_\mathrm{esc,high}$) control the slope and normalization of the power law.  For halos with mass below (above) $M_\mathrm{low}$ ($M_\mathrm{high}$), we set the value of $f_\mathrm{esc}$ to $f_\mathrm{esc,low}$ ($f_\mathrm{esc,high}$). \\

\textbf{MH-Pos}: \cite{Gnedin2008a} and \cite{Wise2009} show that heavy amounts of star formation can create ionized channels within star-forming clouds, providing an easy escape route for ionizing photons.  Star formation is the largest in high-mass galaxies, which tend to live in high-mass halos.  Hence, for this model, we use a prescription wherein the escape fraction scales as a positive power law with halo mass,

\begin{align}
\log_{10} \left(1 -f_\mathrm{esc}\right) =& \log_{10}\left(1 -f_\mathrm{esc,low}\right) - \tag*{} \\ & \left[ \frac{\log_{10}\frac{M_\mathrm{H}}{M_\mathrm{H,low}}}{\log_{10}\frac{M_\mathrm{H}}{M_\mathrm{H,high}}} \log_{10}\frac{1 - f_\mathrm{esc,low}}{1 - f_\mathrm{esc,high}}\right].
\end{align}

\noindent The calibration constants are defined identically to the MH-Neg model.\\

\textbf{Ejected}:  High-resolution hydrodynamical simulations show that feedback is critical in destroying star-forming clouds and allowing easy escape of ionizing photons \citep[e.g.,][]{Kimm2014,Xu2016,Kimm2017}.  Within \texttt{RSAGE}, we capture this by calculating $f_\mathrm{ej}$, the fraction of baryons that have been ejected from the galaxy compared to the number remaining as hot and cold gas.  By setting $f_\mathrm{esc}$ to scale positively with $f_\mathrm{ej}$, we hence allow feedback processes to dictate the instantaneous escape fraction,

\begin{equation}
f_\mathrm{esc} = \alpha f_\mathrm{ej} + \beta.
\end{equation}

\noindent We choose a linear function for simplicity with the strength of coupling controlled by $\alpha$ and the zero-point offset given by $\beta$. \\

\textbf{SFR}: In the MH-Pos model, we use the halo mass as a proxy for the star formation activity that creates ionized channels within the galaxy gas cloud.  As \texttt{RSAGE} explicitly models the evolution of galaxies, this model allows $f_\mathrm{esc}$ to scale explicitly with the galaxy star formation rate (SFR) in the form of a logistic curve,

\begin{equation}
f_\mathrm{esc} = \frac{\delta}{1 + \exp{\left(-\alpha\left(\log_{10}\mathrm{SFR} - \beta\right)\right)}}.
\end{equation}

\noindent This functional form was chosen as $\log_{10}\mathrm{SFR}$ can (theoretically) span $\left[-\infty, +\infty\right]$, aligning itself to the domain of the logistic curve.  Furthermore, the logistic curve has range $[0, \delta]$ allowing easy scaling of the maximum $f_\mathrm{esc}$ value.  Finally, $\alpha$ sets the steepness of the curve and $\beta$ controls the value of $\log_{10}\mathrm{SFR}$ that corresponds to $f_\mathrm{esc} = \delta / 2$.

\subsection{Average Escape Fraction}

\begin{figure*}
	\includegraphics[scale = 0.33]{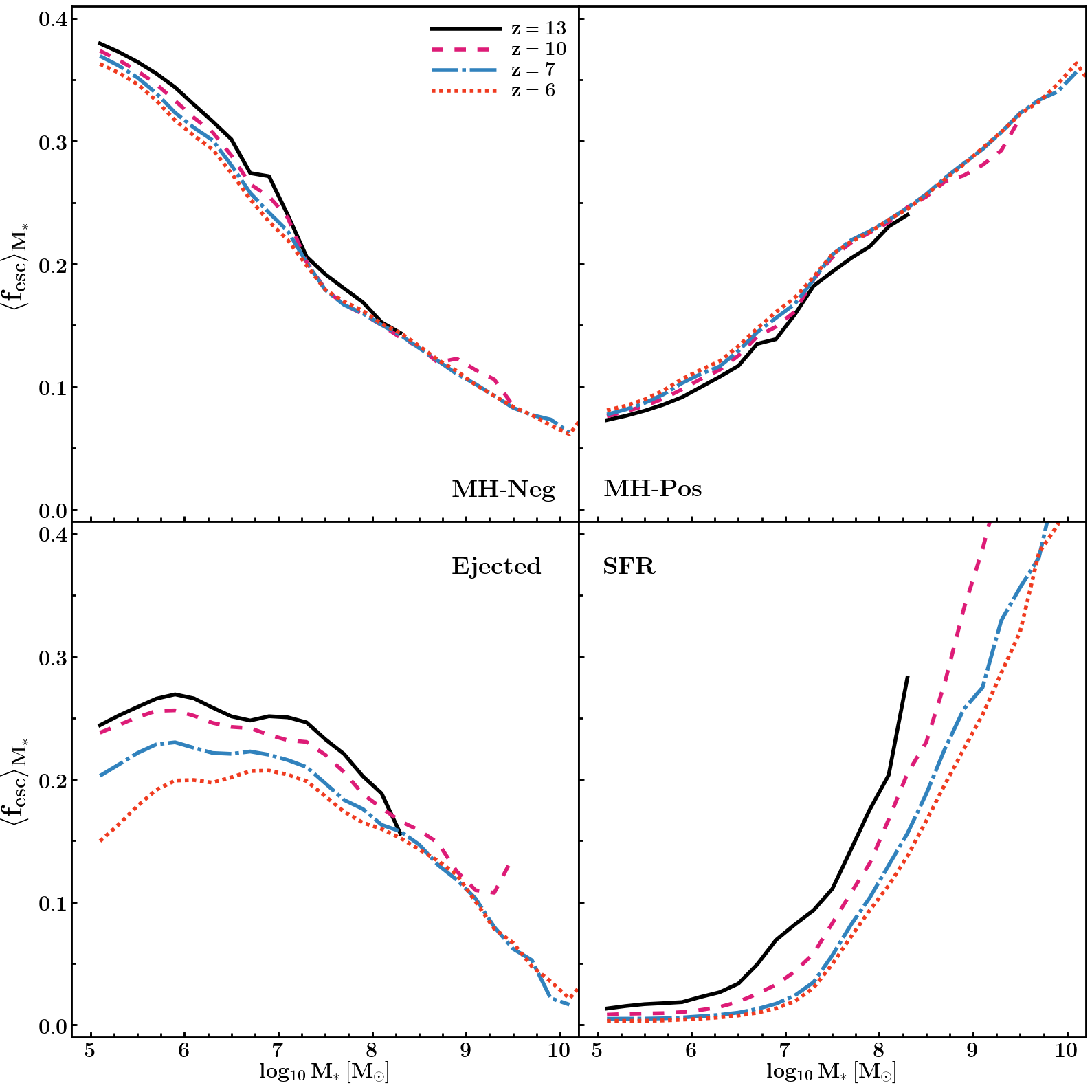}
	\caption{Mean escape fraction within each stellar mass bin for each of the $f_\mathrm{esc}$ models. Calibration parameters (Table \ref{tab:Models}) are chosen to match the inferred estimates of the ionizing emissivity from \protect\cite{Bouwens2015} and measurements of the Thomson optical depth from \protect\cite{Planck2018}.}
	\label{fig:fescAverage}
\end{figure*}

Figure \ref{fig:fescAverage} shows $\langle f_\mathrm{esc}\rangle_{M_*}$, the mean value of $f_\mathrm{esc}$ across all galaxies in a stellar mass bin, as a function of stellar mass. Since high-mass galaxies live within high-mass halos, $\langle f_\mathrm{esc}\rangle_{M_*}$ scales negatively and positively with stellar mass for the MH-Neg and MH-Pos models respectively.  We find a small redshift evolution in $\langle f_\mathrm{esc}\rangle_{M_*}$ at low stellar mass for these two models, resulting from the scatter in the stellar mass-halo mass relationship \cite[see Figure 3 of ][]{Mutch2016}.

Since feedback effects are able to eject baryons more easily within low-mass ($M_* < 10^7 \mathrm{M}_\odot$) galaxies, $\langle f_\mathrm{esc}\rangle_{M_*}$ scales negatively with stellar mass for the Ejected model.  As more massive galaxies tend to have more star formation activity, the SFR model scales positively with stellar mass.

As the redshift decreases, the star formation rate of our simulated galaxies at fixed stellar mass drops over time, in agreement with observations and theory \cite[e.g.,][]{Sparre2015,Santini2017}.  As the ejection of baryonic material is heavily driven through supernovae activity, a drop in the star formation rate will correspond to less supernovae, allowing galaxies to retain more of their baryonic material.  These two phenomena, a decrease in the ejection of baryons and the star formation rate, manifest in our models as $\langle f_\mathrm{esc}\rangle_{M_*}$ decreasing over time for the Ejected and SFR models, respectively.

\subsection{Ionizing Emissivity and Optical Depth} \label{sec:IonizingEmissivity}

\begin{figure*}
	\includegraphics[scale = 0.4]{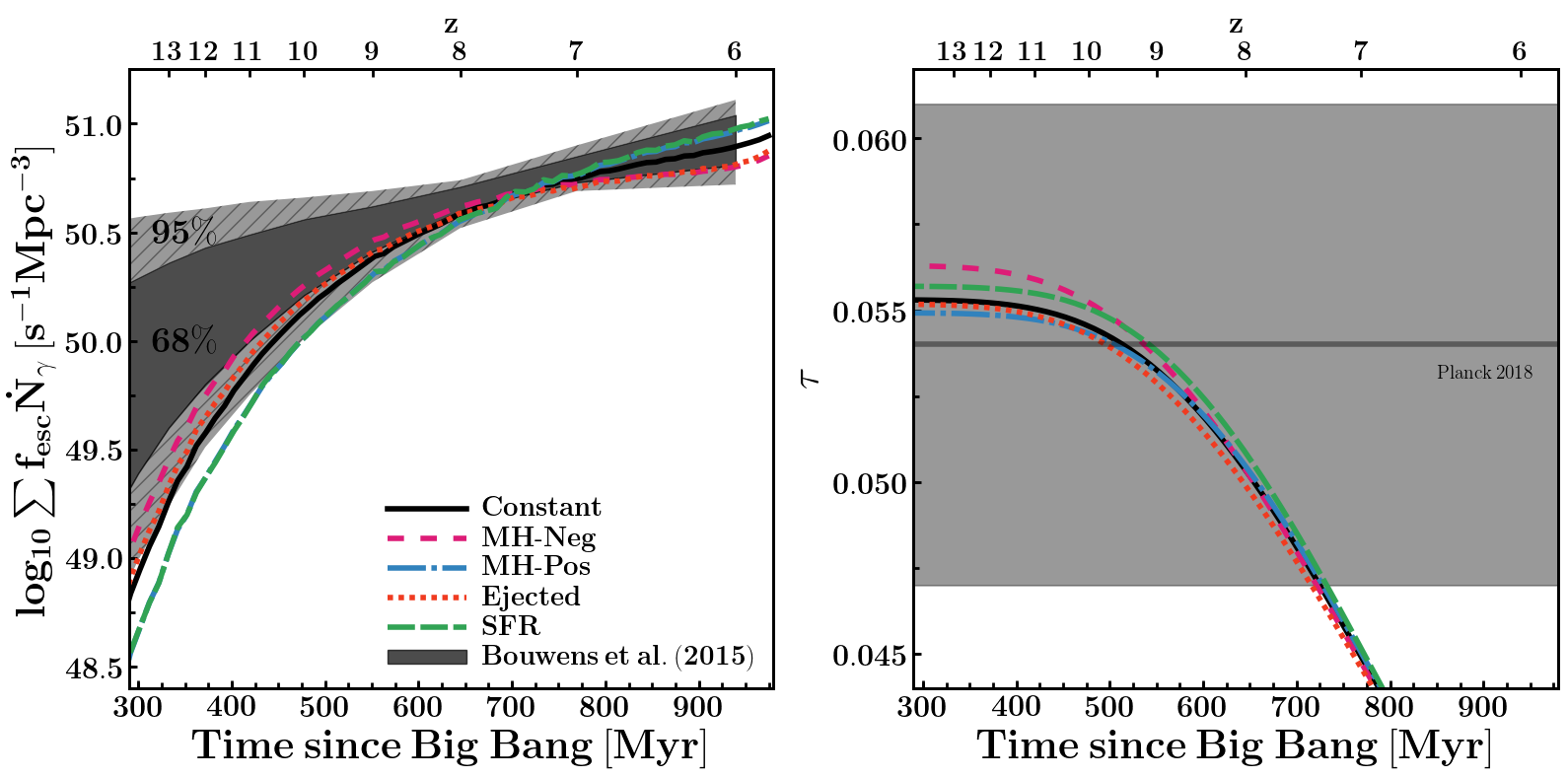}
	\caption{Left: Ionizing emissivity for each of the $f_\mathrm{esc}$ models.  The shaded contours show the derived $68\%$ and $95\%$ confidence intervals for the ionizing emissivity inferred using the Thomson optical depth, quasar absorption spectra, and the prevalence of Ly$\alpha$ emission in redshift $z = 7$ to $z = 8$ galaxies \citep[Table 2 of ][]{Bouwens2015}. Right: Thomson optical depth with the $68\%$ confidence interval measurements of \protect\cite{Planck2018} shown as the shaded region.}
	\label{fig:Nion}
\end{figure*}

The left-hand panel of Figure \ref{fig:Nion} compares the evolution of the ionizing emissivity for each model with the inferred estimates of \cite{Bouwens2015}.  For all models, the general shape and values of the ionizing emissivity match well with observational estimates, an outcome of the free parameters (as shown in Table \ref{tab:Models}) used for each model.  We emphasize that these parameters were chosen to ensure that the more tightly constrained lower redshift ($z=6$ to $z=8$) estimates of \cite{Bouwens2015} were matched closely.

For the majority of our results and calibrations, there is a distinct difference between the $f_\mathrm{esc}$ models that scale negatively with stellar mass (MH-Neg and Ejected) and those that scale positively with stellar mass (MH-Pos and SFR). This distinction is highlighted by the ionizing emissivity in the left-hand panel of Figure \ref{fig:Nion} whose evolution is primarily driven by the growth of the stellar mass function.  At early times (redshift $z > 10$), the stellar mass budget is dominated by low-mass galaxies.  Hence, the MH-Neg and Ejected models will have the largest values of $f_\mathrm{esc}$ and thus ionizing emissivity.  Over time, massive galaxies become more numerous, shifting the stellar mass budget towards the high-mass end.  This results in the $f_\mathrm{esc}$ values of the MH-Pos and SFR models growing quickly, leading to a rapid evolution in the ionizing emissivity.  Eventually, at redshift $z \simeq 7.5$, the rapid growth in $f_\mathrm{esc}$ allows the ionizing emissivity of the MH-Pos and SFR models to surpass all others.  Finally, we see that the ionizing emissivity of the Constant model remains firmly in the middle of the pack throughout all of reionization, a result of its $f_\mathrm{esc}$ values not scaling with stellar mass.

We show the evolution of the Thomson optical depth $\tau$ for each model in the right-hand panel of Figure \ref{fig:Nion} with the measured values of \cite{Planck2018} shown as the shaded region.  All models fall comfortably within the observational constraints, a result of calibrating the models to match the \cite{Bouwens2015} ionizing emissivity which used the optical depth as a key constraint.  Interestingly, we do not find differences between the models depending upon whether $f_\mathrm{esc}$ scales positively or negatively with stellar mass.  This hints that an integrated property such as $\tau$ cannot accurately distinguish between different $f_\mathrm{esc}$ models.  We explore this conclusion in Section \ref{sec:History}.

\section{Reionization History} \label{sec:History}

\begin{figure}
	\includegraphics[scale = 0.4]{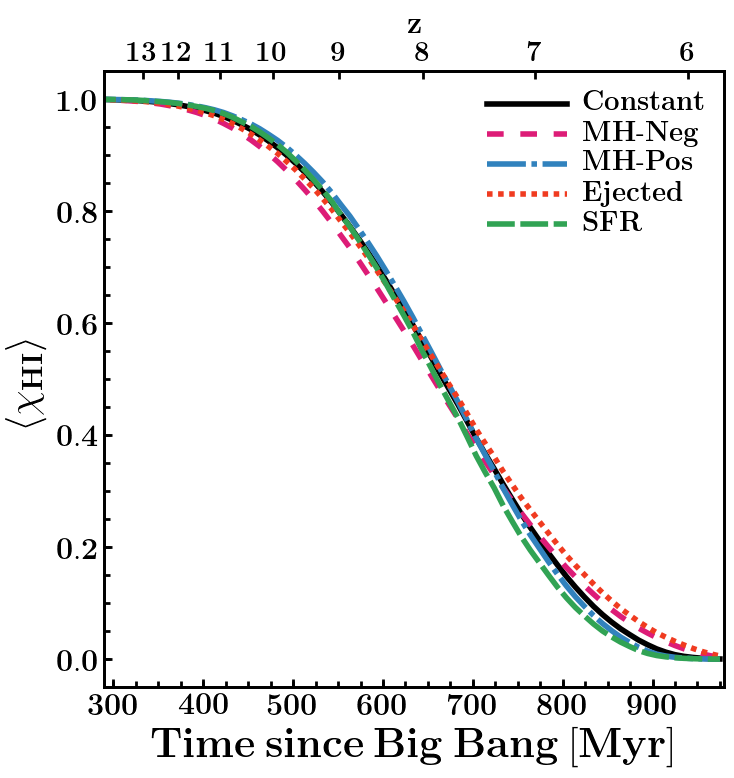}
	\caption{Evolution of the mass-averaged global neutral hydrogen fraction. Despite taking longer to initially begin reionization, the MH-Pos and SFR models rapidly ionize the universe, producing similar durations of reionization (see Table \ref{tab:Duration}).}
	\label{fig:reionhistory}
\end{figure}

We now investigate how the different models of $f_\mathrm{esc}$ affect the global evolution of ionized hydrogen during the Epoch of Reionization. Figure \ref{fig:reionhistory} shows the evolution of the mass-averaged neutral hydrogen fraction, $\langle \chi_{\mathrm{HI}}\rangle$.   We find that $\langle \chi_{\mathrm{HI}}\rangle$ evolves similarly for all models.  Reionization begins slowly at redshift $z \simeq 11$-$12$, corresponding to the appearance of the brightest sources ionizing their immediate surroundings.  As more galaxies are formed, the ionizing emissivity increases and reionization speeds up, highlighted by $\langle \chi_{\mathrm{HI}}\rangle$ becoming steeper from redshifts $z\simeq 10$ to $z \simeq 8$. Below redshift $z \simeq 8$, reionization begins to slow down as the majority of the simulation box is already ionized.

When comparing the evolution of $\langle \chi_{\mathrm{HI}}\rangle$ for different $f_\mathrm{esc}$ models, we find a similar story to that told by the ionizing emissivity in the left-hand panel of Figure \ref{fig:Nion}.  Due to the higher number of ionizing photons at redshift $z \sim 14$, the MH-Neg and Ejected models begin reionization first, reaching $\langle \chi_{\mathrm{HI}}\rangle = 0.99$ at redshift $z = 12.48$ and $z = 12.22$, respectively, compared to the redshift $z = 11.76$ for both the MH-Pos and SFR models.  However, despite the MH-Neg and Ejected models starting reionization earliest, they do not finish first. Rather, the MH-Pos and SFR models reionize the universe the quickest and by redshift $z \simeq 8$ these two models outpace all others and finish reionization sooner.  This echoes the results of the previous section where we found that the ionizing emissivity of the MH-Pos and SFR models grow the quickest.  Once again, we see that the Constant model divides these two regimes.

This difference in reionization speed is summarized in Table \ref{tab:Duration} where we list the duration\footnote{Here we use the definition of ``duration'' being the time between $\langle \chi_{\mathrm{HI}}\rangle = 0.90$ ($z_{90\%}$) and $\langle \chi_{\mathrm{HI}}\rangle = 0.01$ ($z_{1\%}$).} of reionization for each model.  We find that despite the MH-Neg and Ejected models starting their reionization of the universe first, they have a slower, more extended reionization history compared to the other models.  Indeed, the rapid ionizing emissivity evolution in the MH-Pos and SFR models results in a quick Epoch of Reionization.  This highlights that, whilst a deficiency of ionizing photons at very early times (redshift $z \sim 14$) leads to a delayed start of reionization, the overall duration of reionization is heavily controlled by the growth of the ionizing emissivity.

\begin{table}
	\centering
	\begin{tabular}{|c|c|c|c|c|}
		\hline
		\textbf{Model} & $\mathbf{z_\mathrm{90\%}}$ & $\mathbf{z_\mathrm{50\%}}$ & $\mathbf{z_\mathrm{1\%}}$ & $\mathbf{\Delta z \left(\Delta t \: [{Myr}]\right)}$\\ \hline \hline
		Constant & 9.77 & 7.83 & 6.08 & 3.70 (432.0) \\ \hline
		MH-Neg  & 10.07 & 7.92 & 5.92 & 4.15 (482.2) \\ \hline
		MH-Pos  & 9.63 & 7.44 & 6.18 & 3.45 (401.8) \\ \hline
		Ejected & 9.92 & 7.74 & 5.88 & 4.04 (482.2) \\ \hline
		SFR & 9.77 & 7.83 & 6.23 & 3.54 (401.8) \\ \hline
	\end{tabular}
	\caption{Duration of reionization for each of the $f_\mathrm{esc}$ models.  $z_\mathrm{90\%}$, $z_\mathrm{50\%}$ and $z_\mathrm{1\%}$ denote the redshifts at which the universe is $90\%$, $50\%$ and $1\%$ neutral with $\Delta z = z_\mathrm{90\%} - z_\mathrm{1\%}$ (also shown in Myr for clarity).} 
	\label{tab:Duration}
\end{table}

The core motivation of this work is to investigate the possibility of distinguishing between different physically motivated models of $f_\mathrm{esc}$.  From Table \ref{tab:Duration}, we find that an $f_\mathrm{esc}$ model that scales negatively with stellar mass (Ejected) takes $\simeq$$80$Myr longer to complete reionization than a model that scales positively with stellar mass (SFR).  This story remains the same regardless of the exact definition of `duration' used.  Using another common definition of $\Delta z = z_\mathrm{80\%} - z_\mathrm{1\%}$ \citep[e.g.,][]{Zahn2012,George2015}, we again find that the SFR model finishes reionization $\simeq 80$Myr earlier than all other models.  To check the robustness of this difference in duration, we must consider how an uncertainty in $\tau$ affects our result.   We focus on the Ejected model and adjust the free parameter $\alpha$ from $0.30$ to $0.45$, increasing the value of $\tau$ from $0.054$ to $0.061$\footnote{$\tau = 0.061$ corresponds to the largest $\tau$ value that remains consistent with \cite{Planck2018}}.  This recalibration results in a duration of reionization of $391.8$Myr, below the fiducial Constant model. Hence, with the current uncertainty in $\tau$, the duration of reionization could not differentiate between a Constant $f_\mathrm{esc}$ model and one that scales positively with stellar mass (e.g., the SFR model) and has a higher value of $\tau$.  We conclude that, with the current uncertainty in $\tau$, the duration of reionization cannot be used alone to differentiate between the different $f_\mathrm{esc}$ models.

\section{The Topology of Reionization} \label{sec:Morphology}

From Section \ref{sec:History}, we found that an integrated property, such as the optical depth, could not differentiate between the $f_\mathrm{esc}$ models.  In this section, we analyse the topology of ionized regions during Reionization.  Specifically, we first investigate how the spatial topology of the ionized hydrogen differs between each model.  We then quantify these differences using the $21$cm power spectrum and comment on the possibility of the upcoming Square Kilometre Array (SKA) distinguishing between the models.

\subsection*{Spatial Slices Through the Ionization Field}

\begin{figure*}
	\hspace*{-0.4in}
	\includegraphics[scale = 0.65]{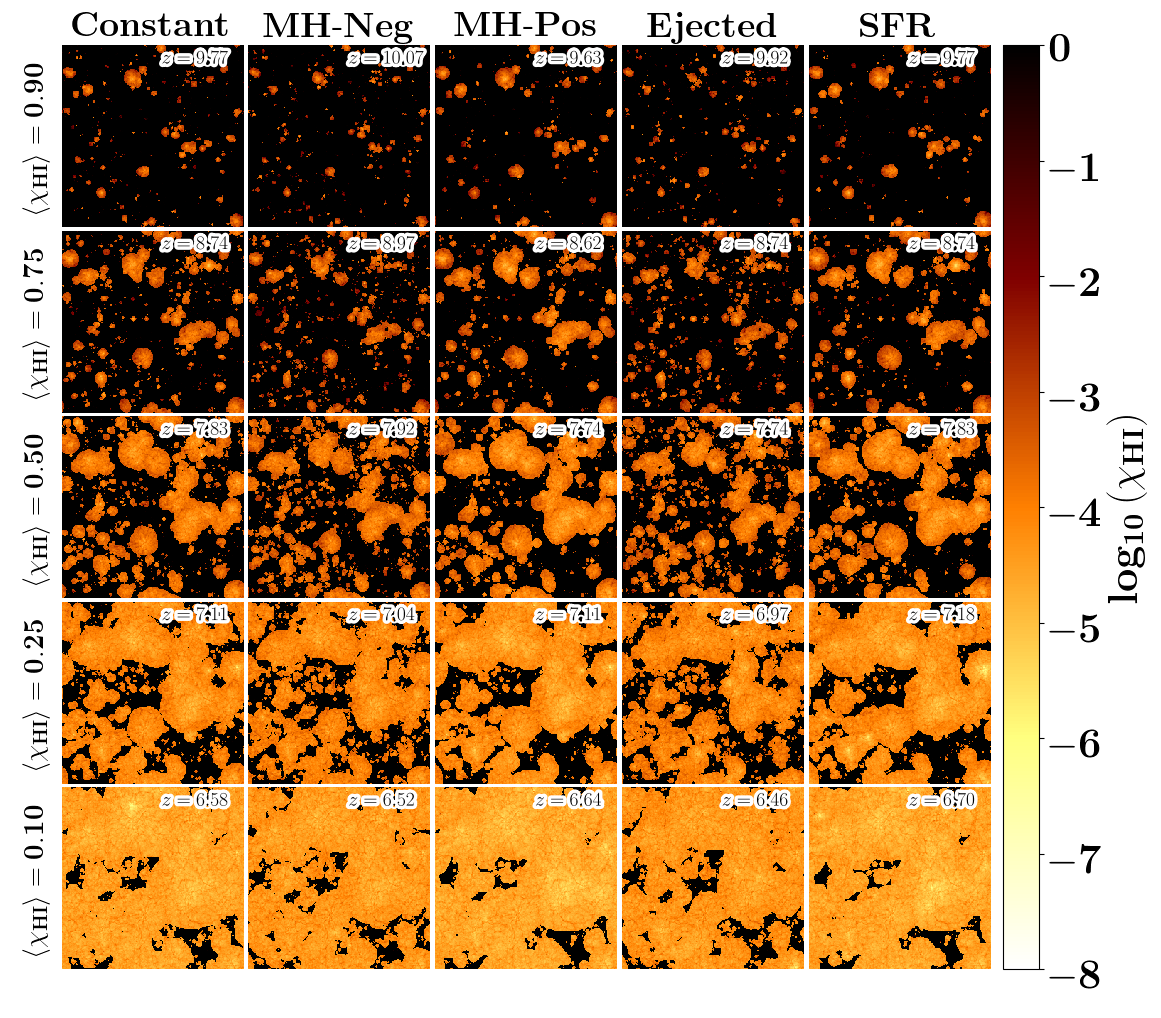}
	\vspace*{-0mm} 
	\caption{Slices through the ionization field for each $f_\mathrm{esc}$ model (columns) at various fixed global neutral hydrogen fractions (rows).  Slices are one grid cell thick ($0.39\mathrm{h}^{-1}$Mpc).  The colourbar shows the local ionization fraction within each grid cell with black corresponding completely neutral and white denoting (almost) complete ionization.  For the models that scale positively (negatively) with stellar mass we have a low (high) number of large (small) ionized regions scattered throughout the box.}
	\label{fig:slices} 
\end{figure*}

To investigate differences in the spatial topology of ionized hydrogen, in Figure \ref{fig:slices} we show one grid cell ($0.39h^{-1}$Mpc) slices through the ionization field for each $f_\mathrm{esc}$ model at fixed global neutral hydrogen fractions\footnote{We compare at fixed neutral hydrogen fractions to ensure we are comparing the same amount of ionization.}. We remind the reader that from Figure \ref{fig:StellarMass}, we have orders of magnitude more low-mass ($\mathrm{M}_* \leq 10^{8}\mathrm{M}_\odot$) galaxies compared to high-mass ($\mathrm{M}_* \geq 10^{9}\mathrm{M}_\odot$) ones.  Due to the relatively low star formation rate of these low-mass galaxies, they will produce fewer ionizing photons.  Hence for all models, we expect a large number of low-powered ionizing sources in conjunction with a handful of massive objects that emit a large number of ionizing photons.  This is evident in the Constant model, where we have numerous small ionized regions scattered throughout the simulation box alongside a handful of large, extended ionized regions.

For the other models, we must consider how $f_\mathrm{esc}$ scales with galaxy stellar mass (Figure \ref{fig:fescAverage}).  For the MH-Neg and Ejected models, $f_\mathrm{esc}$ is largest for low-mass galaxies.  Hence, compared to the Constant model at a fixed neutral hydrogen fraction, the number of small ionized regions increases whilst the number of large ionized regions decreases.  We quantify this in Table \ref{tab:BubbleSize} where we show the mean size of ionized regions for the models. Due to the higher number of small ionized regions, the MH-Neg and Ejected models have a smaller mean size compared to the constant model: $11.85h^{-1}$Mpc and $11.79h^{-1}$Mpc compared to $13.93h^{-1}$Mpc at $\langle \chi_{\mathrm{HI}}\rangle = 0.25$. Conversely for the MH-Pos and SFR models, $f_\mathrm{esc}$ scales positively with stellar mass and this reasoning is reversed: The number of small ionized regions is suppressed whilst the number of large ionized regions is enhanced. This is shown again in Table \ref{tab:BubbleSize} where the mean size of ionized regions is larger for MH-Pos and SFR models compared to the Constant model: $17.23h^{-1}$Mpc and $18.58h^{-1}$Mpc compared to $13.93h^{-1}$Mpc at $\langle \chi_{\mathrm{HI}}\rangle = 0.25$.

\begin{table}
	\centering
	\begin{tabular}{|c|c|c|c|}
		\hline
		& \multicolumn{3}{|c|}{$\mathbf{Mean \: Region \: Size \: [h^{-1}Mpc]}$} \\ \hline
		\textbf{Model} & $\mathbf{\langle\chi_{\mathrm{HI}} = 0.75\rangle}$ & $\mathbf{\langle\chi_{\mathrm{HI}} = 0.50\rangle}$ & $\mathbf{\langle\chi_{\mathrm{HI}} = 0.25\rangle}$\\ \hline \hline
		Constant & 3.10 & 6.14 & 13.93 \\ \hline
		MH-Neg  & 2.34 & 4.67 & 11.85 \\ \hline
		MH-Pos  & 3.82 & 7.62 & 17.23 \\ \hline
		Ejected & 2.61 & 4.97 & 11.79 \\ \hline
		SFR & 4.19 & 8.19 & 18.58 \\ \hline
	\end{tabular}
	\caption{Mean size of ionized regions for each model at fixed neutral hydrogen fractions.  To calculate this we first mark any cell that has ionization fraction $\chi_{\mathrm{HI}} > 0.9$ as `ionized' and select a random ionized cell.  Then we walk in a random axis-aligned direction (i.e., either $\pm x$, $\pm y$ or $\pm z$) and count the number of cells until we reach a neutral cell.  We repeat this process $10\: 000$ times to determine the representative size of ionized regions for each $f_\mathrm{esc}$ model.} 
	\label{tab:BubbleSize}
\end{table}

The ionized region morphology matches the results of \cite{McQuinn2007}, who find that the regions grow larger as the ionizing sources become rarer; that is, the mean region size increases as high-mass sources dominate the ionizing photon budget.  A similar conclusion is also made by \cite{Greig2015}.  Finally, \cite{Geil2016} utilize the \texttt{Meraxes} model of \cite{Mutch2016} and implement a scenario in which only galaxies hosted by high-mass ($M_\mathrm{H} > 10^{10}\mathrm{M}_\odot$) halos contribute to reionization.  Under this condition, they find that the average size of ionized regions increases compared to a fiducial model which allows contribution from all galaxies.  This aligns with our results in Table \ref{tab:BubbleSize}, where the MH-Pos model contains the largest ionized regions.

\subsection*{21cm Power Spectrum}

To observationally map the size distribution of the ionization field, one common technique is to use the $21$cm hydrogen emission line.  This signal is extremely sensitive to the presence of neutral hydrogen, providing the perfect tool for constraining the history, topology, and sources of reionization. In the remainder of this section, we focus on the signal's ability to differentiate between the topologies of the $f_\mathrm{esc}$ models.

The $21$cm differential brightness temperature depends on fluctuations in both the ionization and dark matter density fields and is calculated for each cell in the simulation box as \citep{Iliev2012},

\begin{equation}
\delta T_\mathrm{b}\left(\mathbf{x},z\right) = T_0\left(z\right) \chi_{\mathrm{HI}}\left(\mathbf{x},z\right) \delta \left(\mathbf{x},z\right) \: \mathrm{mK},
\label{eq:Tb}
\end{equation}

\noindent with

\begin{equation}
T_0\left(z\right) = 28.5 \left(\frac{1+z}{10}\frac{0.15}{\Omega_mh^2}\right)^{1/2} \left(\frac{\Omega_bh^2}{0.023}\right),
\end{equation}

\noindent where $\chi_\mathrm{HI}\left(\mathbf{x},z\right)$ is the neutral hydrogen fraction within each grid cell, $\delta \left(\mathbf{x},z\right)$ is the dark matter overdensity defined as $\delta = \rho/\langle \rho \rangle$ with density $\rho$, $h$ is the Hubble parameter, and $\Omega_m$ and $\Omega_b$ denote the critical cosmological matter and baryonic densities, respectively.  To calculate the dimensional $21$cm power spectrum, $\Delta_{21}^2\left(k,z\right)$, we define $\Delta \widetilde{T}_\mathrm{b}\left(\mathbf{k},z\right)$ as the Fourier transform of Equation \ref{eq:Tb} with $\mathbf{k} = \left(k_x, k_y, k_z\right)$ denoting the 3-dimensional wavenumber,

\begin{equation}
\Delta_{21}^2\left(k,z\right) = 4\pi k^3 \langle \Delta \widetilde{T}_\mathrm{b}\left(\mathbf{k},z\right)\Delta \widetilde{T}_\mathrm{b}\left(-\mathbf{k},z\right)\rangle \: \mathrm{mK}^2,
\end{equation}

\noindent where $\langle \rangle$ denotes the spherically averaged value and we use the \texttt{numpy} Python package \citep{numpy} to compute the Fourier transform which dictates the pre-factor value of $4\pi k ^3$. Due to the numerical resolution of our simulations and its impact on the power spectrum, we limit our analysis to scales $k < 4.0h\:\mathrm{Mpc}^{-1}$.  

\begin{figure*}
	\includegraphics[scale = 0.4]{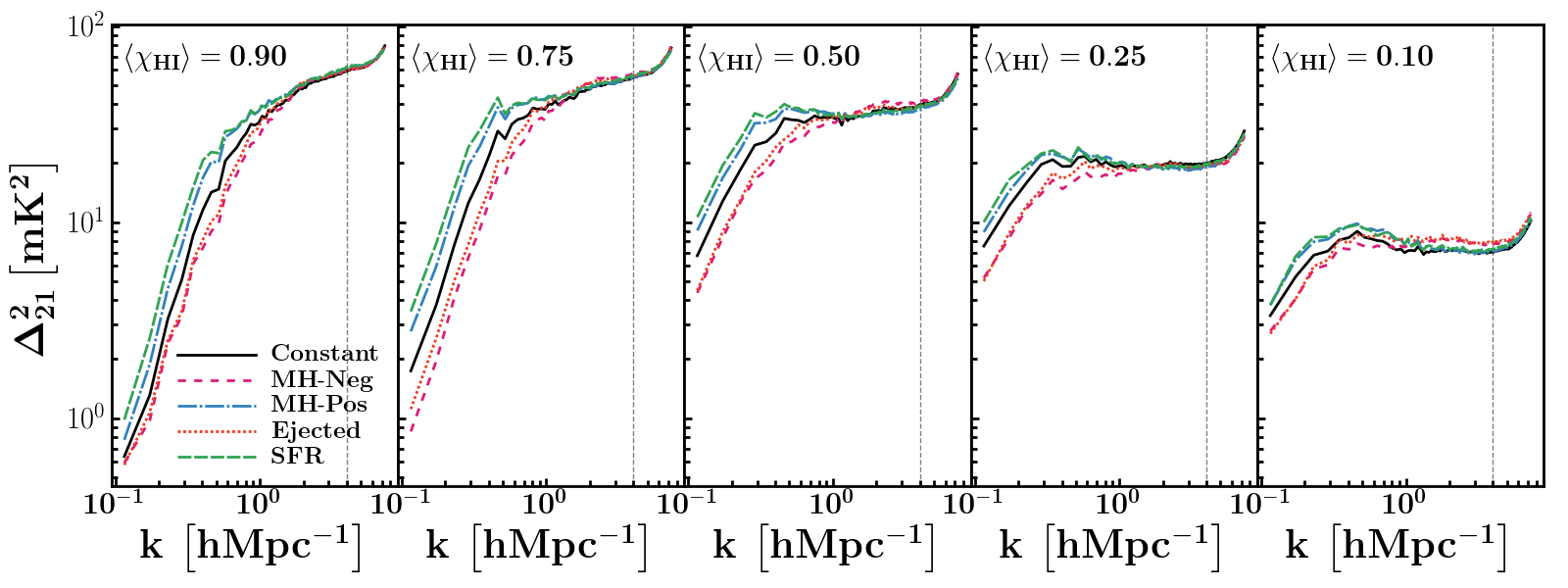}
	\caption{21cm Power Spectra for all $f_\mathrm{esc}$ models at fixed neutral hydrogen fractions. Due to the resolution of our simulation, we limit analysis to wavenumbers below $k < 4.0h\mathrm{Mpc}^{-1}$, shown as the vertical dotted line.}
	\label{fig:21cm}
\end{figure*}

\begin{figure}
	\includegraphics[scale = 0.4]{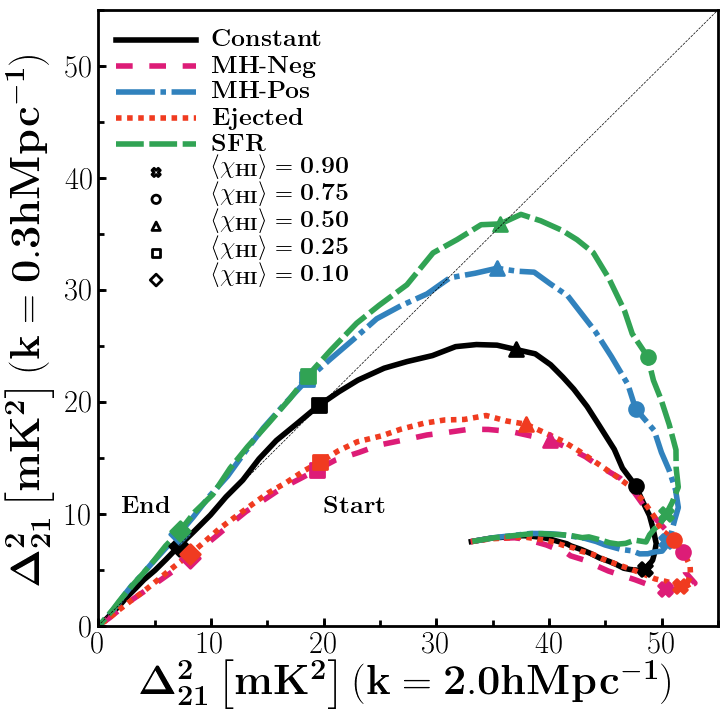}
	\caption{Evolution of $21$cm large-scale ($k = 0.3h \mathrm{Mpc}^{-1}$) power as a function small-scale ($k = 2.0h \mathrm{Mpc}^{-1}$) power.  The global neutral fractions marked correspond to the full spectra shown in Figure \ref{fig:21cm}.  The thin black dashed line denotes the one-to-one line; below (above) this line we are small (large) scale dominated.}
	\label{fig:ScalevScale}
\end{figure}

We first discuss the general evolution and features of the $21$cm power spectrum.  In Figure \ref{fig:21cm} we show the $21$cm power spectrum at fixed neutral hydrogen fractions. To better understand the full redshift evolution of the spectrum, its evolution at specific scales, and the topological differences between models, we also show the large-scale power as a function of small-scale power (hereafter called `scale space') in Figure \ref{fig:ScalevScale}.  Here we define `large-scale' as $k = 0.3h \mathrm{Mpc}^{-1}$ and `small-scale' as $k = 2.0 h\mathrm{Mpc}^{-1}$.

At the beginning of reionization, when the universe is mostly neutral, the variance in $\delta T_\mathrm{b}$ (and hence $21$cm power) is driven by dark matter fluctuations, causing the $21$cm power to follow the underlying density distribution. As the first small ionized regions begin to appear, all models show an increase in small-scale power.  Furthermore, \cite{Lidz2008} highlight that this period also corresponds to an `equilibration' phase where overdense and underdense regions have similar brightness temperatures. As a result, this equilibration phase also results in a decrease in large-scale power from the start of our simulation to $\langle \chi_\mathrm{HI} \rangle = 0.90$.  This is shown in Figure \ref{fig:ScalevScale} where we see an increase in small-scale power at the expense of large-scale power.

As reionization progresses from $\langle \chi_\mathrm{HI} \rangle = 0.90$ to $\langle \chi_\mathrm{HI} \rangle = 0.50$, small, isolated ionized regions grow and eventually begin to overlap, reducing small-scale power whilst boosting it on large scales. Finally, as reionization passes its midpoint, the majority of the simulation cells are ionized.  Consequently, beyond $\langle \chi_\mathrm{HI} \rangle = 0.50$, the $21$cm brightness temperature and the resulting power decreases on all scales. This is highlighted in Figure \ref{fig:ScalevScale} where the turning point for all models lies close to $\langle \chi_\mathrm{HI} \rangle = 0.50$.

We now comment on the differences in the $21$cm power spectra between the $f_\mathrm{esc}$ models.  Since reionization begins at different times for each model (Table \ref{tab:Duration}), the redshift, and hence dark matter density, at $\langle \chi_{\mathrm{HI}}\rangle = 0.90$ is different.  As mentioned above, the $21$cm power during the beginning of reionization is dominated by fluctuations in the dark matter density field.  Furthermore, as the MH-Neg and Ejected models reach $\langle \chi_{\mathrm{HI}}\rangle = 0.90$ at the highest redshift, their dark matter fluctuations are the smallest on large scale and hence they initially have the smallest $21$cm power on large scales. By the same logic, the MH-Pos and SFR models have the largest $21$cm power on large scales at this time, shown most clearly in the left-most panel of Figure \ref{fig:21cm}.

From Table \ref{tab:BubbleSize}, we saw that, compared to the Constant model, the MH-Neg and Ejected models have smaller ionized regions whereas the MH-Pos and SFR models have larger ionized regions during the intermediate ($\langle \chi_{\mathrm{HI}}\rangle = 0.25$, $\langle \chi_{\mathrm{HI}}\rangle = 0.50$ and $\langle \chi_{\mathrm{HI}}\rangle = 0.75$) stages of reionization.  This is reflected in the $21$cm power spectrum where we find increased small-scale power for the MH-Neg and Ejected models and enhanced large-scale power for the MH-Pos and SFR models. This is a key finding of our work: Allowing $f_\mathrm{esc}$ to scale negatively (positively) with stellar mass increases power on small (large) scales. Figure \ref{fig:ScalevScale} highlights this where we see a marked difference in the large-scale power across all models. In particular, we see that the MH-Neg and Ejected models never have more large-scale power than small-scale, providing a powerful diagnostic for models of $f_\mathrm{esc}$ that scale negatively with stellar mass.

Overall, the behaviour of the $21$cm power spectra across our different $f_\mathrm{esc}$ models matches the general trends found in the literature.  In particular, \cite{Dixon2016} and \cite{Mesinger2016} find that aggressively suppressing low-mass sources, analogous to allowing $f_\mathrm{esc}$ to scale with stellar mass, enhances large-scale $21$cm power. Finally, \cite{Kim2013} also find that implementing an $f_\mathrm{esc}$ that increases with halo mass reduces $21$cm power on scales $k < 0.4$ hMpc$^{-1}$, which is identical to the behaviour of our MH-Pos model.

\subsection*{Distinguishing Models with SKA}

We now focus on the possibility of using the $21$cm power spectrum to observationally distinguish between different models of $f_\mathrm{esc}$.  Specifically, we focus on the Square Kilometre Array Low Frequency instrument (SKA1-Low). 

\begin{figure*}
	\includegraphics[scale = 0.4]{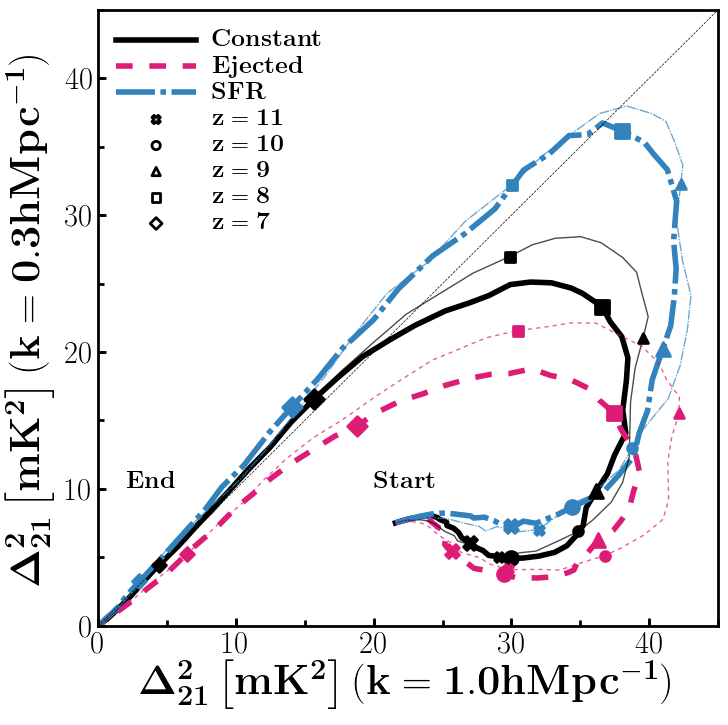}
	\includegraphics[scale = 0.4]{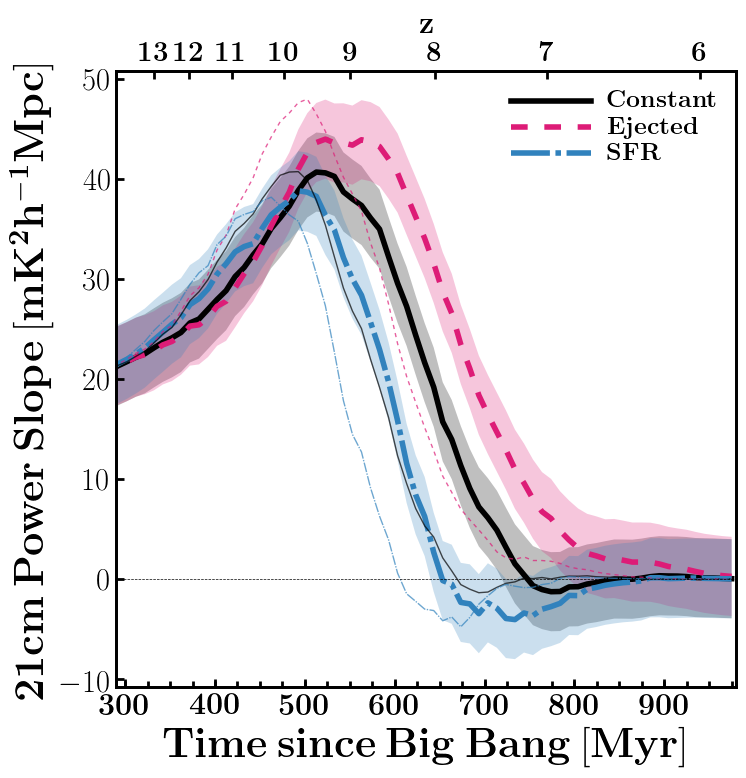}
	\caption{Left: Evolution of $21$cm large-scale ($k = 0.3h \mathrm{Mpc}^{-1}$) power as a function small-scale ($k = 1.0h \mathrm{Mpc}^{-1}$) power.  The thin black dashed line denotes the one-to-one line; below (above) this line we are small (large) scale dominated.  Right: The evolution of the $21$cm power spectrum slope between large and small scales.  The shaded regions show the SKA sensitivity assuming $200$ hours integration time over a $10$MHz bandwidth.  For both panels, thick lines show fiducially calibrated models; thin lines are the result of rescaling the calibration parameters such that each model has $\tau = 0.061$.}
	\label{fig:ObsScaleSpace}
\end{figure*}

In Figure \ref{fig:ScalevScale}, we chose the scales to best elucidate the difference in trajectories for each $f_\mathrm{esc}$ model.  Whilst $k = 0.3h \mathrm{Mpc}^{-1}$ aligns with the large scales probed by the SKA1-Low,  we show these trajectories for a more attainable small-scale wavenumber $k = 1.0h \mathrm{Mpc}^{-1}$ in the left-hand panel of Figure \ref{fig:ObsScaleSpace}.  At this reduced scale, the models do not show significant differences in small-scale power.  Nevertheless, the large-scale power is still noticeably different between $f_\mathrm{esc}$ models that scale positively and negatively with stellar mass.

From the scale-space trajectories in Figure \ref{fig:ObsScaleSpace}, it is difficult to assess the relative large- and small-scale power across models at a fixed redshift.  This will be critical for upcoming radio telescopes which will target specific redshift windows.  To this end, we remove a dimension of the scale-space trajectories and calculate the slope of the $21$cm power spectrum between small and large scales, 

\begin{equation}
m = \frac{\left(\Delta_\mathrm{21,large}^2 - \Delta_\mathrm{21,small}^2\right)}{\left(k_\mathrm{21,large} - k_\mathrm{21,small}\right)}.
\label{eq:Slope}
\end{equation}

We show the evolution of the $21$cm slope in the right-hand panel of Figure \ref{fig:ObsScaleSpace}.  Initially, as small-scale power increases at the expense of large-scale power, the slope of the spectrum increases for all models.  Then, as this equilibration phase ends and ionized regions begin to grow, large-scale power grows quickly, as shown by the scale-space trajectories in the left-hand panel of Figure \ref{fig:ObsScaleSpace}, leading to a decline in the slope. After each model reaches the mid-point of reionization, the slope continues to decrease but at a slower rate, mirroring the extended second half of reionization we saw in Figure \ref{fig:reionhistory}. Finally, we see that the large-scale power never exceeds the small-scale power in the Ejected model; hence, the power spectrum slope is never negative for this model.

To make accurate conclusions regarding SKA1-Low's ability to differentiate between the models, we must account for the observational uncertainty associated with the instrument. Using the V4A\footnote{\texttt{http://astronomers.skatelescope.org/wp-content/uploads/\\*2015/11/SKA1-Low-Configuration\_V4a.pdf}} array configuration for the SKA1-Low and matching the system temperature and effective collecting area as a function of frequency to the {\it SKA1 System Baseline Design} document\footnote{\texttt{http://astronomers.skatelescope.org/wp-content/uploads/\\*2016/05/SKA-TEL-SKO-0000002\_03\_SKA1SystemBaselineDesignV2.pdf}}, we calculate the $21$cm power spectrum sensitivity assuming a $10$MHz bandwidth, an integration time of $200$ hours\footnote{We find that $200$ hours of integration is the minimum time required to ensure no overlap of the error bars between redshift $z=7.5-8.5$ in the right panel of Figure \ref{fig:ObsScaleSpace}.} and a redshift window of $z=8$ to $z = 10$. At wavenumbers $k = 0.3h \mathrm{Mpc}^{-1}$ and $k = 1.0h \mathrm{Mpc}^{-1}$, we find an uncertainty of $1.15 \times 10^{-2}\mathrm{mK}^2$ and $1.37\mathrm{mK}^2$, respectively.  Propagating these uncertainties in Equation \ref{eq:Slope}, we show the SKA1-Low $21$cm power spectrum slope uncertainty in the right panel of Figure \ref{fig:ObsScaleSpace} as shaded regions.  We find that above redshift $z \simeq 8.5$ and below redshift $z \simeq 7.5$ the models are indistinguishable.  For $z > 8.5$, reionization has only just begun and the difference in topology has not fully manifested.  Conversely, for $z < 7.5$, reionization is reaching its conclusion where the ionized regions merge and hide topological differences. %Between redshift $z =7-9$ we find that the SKA is able to discriminate between the different models of $f_\mathrm{esc}$.

Finally, we discuss how an uncertainty in the \cite{Planck2018} measurement of $\tau$ affects our result.  Similar to the reionization history, we investigate this by recalibrating our models to produce the largest value of $\tau$ that remains in agreement with \cite{Planck2018} (i.e., $\tau = 0.061$) and show the scale-dependent evolution of the $21$cm power spectrum as thin lines in Figure \ref{fig:ObsScaleSpace}.  Immediately we see that even with an adjusted value of $\tau$, the Ejected model never has more large-scale power than small-scale.  This provides a `smoking gun' to rule out $f_\mathrm{esc}$ models that scale negatively with stellar mass. We find that the increased $\tau$ models exhibit a vertical offset in $m$.  For example, the right-hand panel of Figure \ref{fig:ObsScaleSpace} shows that the increased $\tau$ Constant model is observationally indistinguishable from the fiducial SFR model.  However, we also see that the shape of the scale-space trajectories and $m$ remains largely unchanged for the increased $\tau$ models. Hence, one method to observationally distinguish between models of $f_\mathrm{esc}$ with an uncertain value of $\tau$ is to measure the maximum or minimum values of $m$.  This will require multiple redshift measurements of the $21$cm power spectrum to determine where the maxima/minima lie.  Furthermore, to ensure that the models do not have overlapping observational uncertainties at these points, the integration time must be increased to approximately $800$ hours.

\section{Discussion} \label{sec:Discussion}

%\subsection{Potential Caveats}

For all our models, we did not account for the contribution of quasars to the ionizing emissivity.  In theory, the hard ionizing radiation of quasars would create large ionized regions enhancing $21$cm power on large scales. \cite{Datta2012} find that the size of such regions is comparable to the regions surrounding clustered stellar sources, hinting that the inclusion of quasar radiation could impact the scale-space trajectories (Figure \ref{fig:ScalevScale} and the left-hand panel of Figure \ref{fig:ObsScaleSpace}) by shifting all paths towards the MH-Pos and SFR models.  However, the exact contribution of quasars to the ionizing photon budget remains a contentious topic \citep[e.g.,][]{Kollmeirer2014,Madau2017,Qin2017,Parsa2018}.

A key aspect of reionization simulations is the ability to appropriately model low-mass objects during the earliest stages of the Universe.  In particular, works such as \cite{Choudhury2007}, \cite{Yajima2011}, and \cite{Paardekooper2015} find that dwarf galaxies and mini-halos contribute significantly to the ionizing photons budget only at redshifts $z > 11$. In our work, we do not expect the limited mass resolution of \textit{Kali} (resolving halos of mass $4 \times 10^8\mathrm{M}_\odot$) to significantly affect our results.  Due to radiative feedback, the low-mass galaxies hosted by halos smaller than the resolution limit would have their star formation rapidly quenched, mitigating their overall contribution to the ionizing photon budget.  Furthermore, from Figure \ref{fig:ObsScaleSpace}, the scale-space trajectories at redshifts $z > 11$ are observationally indistinguishable, highlighting the negligible impact of dwarf galaxies on our main findings.

The functional forms of $f_\mathrm{esc}$ for the Ejected and SFR models were chosen to capture the underlying mechanism that links $f_\mathrm{esc}$ to galaxy-scale processes, as highlighted by authors such as \cite{Gnedin2008}, \cite{Wise2009}, \cite{Kimm2014}, \cite{Kimm2017}, and \cite{Trebitsch2017}. However, we do acknowledge that we chose such functions (i.e., linear and logistic) primarily for their simplicity. Whilst selecting different functional forms may produce slightly different results, we stress that the role of these functions were to provide $f_\mathrm{esc}$ models that scale negatively/positively with stellar mass. Importantly, the \texttt{RSAGE} model is constructed to allow any arbitrary form of $f_\mathrm{esc}$, providing a powerful avenue for exploring more complex functional forms \cite[e.g.,][]{Paardekooper2015} as our ability to model and observe $f_\mathrm{esc}$ becomes more nuanced and extensive.  

\section{Conclusion} \label{sec:Conclusion}

In this work we have introduced \texttt{RSAGE}, a new open source\footnote{https://github.com/jacobseiler/rsage} galaxy evolution model that self-consistently accounts for feedback effects associated with the Epoch of Reionization.  Motivated by work highlighting the importance of galaxy-scale feedback on the instantaneous value of $f_\mathrm{esc}$ \citep[e.g.,][]{Paardekooper2015,Kimm2017,Trebitsch2017}, we use the galaxies from our model as ionizing sources and generate unique $f_\mathrm{esc}$ values based on the host galaxy properties.  In particular, we use galaxy feedback and star formation activity to create $f_\mathrm{esc}$ models that scale negatively or positively with stellar mass.  By following the evolution of ionized hydrogen within each model, we assess the possibility of distinguish between different $f_\mathrm{esc}$ models using the duration of reionization or topology of ionized regions.

We find adopting an $f_\mathrm{esc}$ model that scales negatively with stellar mass causes reionization to start early due to the presence of many low-mass galaxies.  However, as galaxies grow more massive over time, the mass-averaged value of $f_\mathrm{esc}$ drops for these models, leading to a slow, extended Epoch of Reionization.  As a result, models that scale positively with stellar mass complete reionization sooner, despite starting much later, due to the comparatively rapid growth in ionizing emissivity. 
Regardless of these different reionization histories, we find that measurements of integrated quantities, such as the optical depth by \cite{Planck2018}, can not distinguish between different $f_\mathrm{esc}$ models.

However, the different $f_\mathrm{esc}$ models leave distinct signatures in the ionization topology. Due to the high number of low-mass galaxies, an $f_\mathrm{esc}$ model that scales negatively with stellar mass will have many galaxies that ionize their immediate surroundings.  This leads to a high number of small ionized regions scattered throughout the simulation box.  Conversely, for an $f_\mathrm{esc}$ model that scales positively with stellar mass, we find (at a fixed global neutral hydrogen fraction) a handful of very large ionized regions.  The constant $f_\mathrm{esc}$ case divides these two regimes.

These differences in the ionization topologies manifest in the power spectra of the $21$cm signal. Since the abundances of the smallest and largest ionized regions represent the key differences in the different $f_\mathrm{esc}$ models, we have plotted the evolution of the large-scale power as a function of the small-scale power. We find that our adopted $f_\mathrm{esc}$ models have distinctly different trajectories through this scale space: Large-scale power never exceeds small-scale power in $f_\mathrm{esc}$ models scaling negatively with stellar mass, while it surpasses small-scale power for $\langle\chi_\mathrm{HI}\rangle\lesssim 0.5$ in $f_\mathrm{esc}$ models scaling positively with stellar mass.

With the relation between the large- and small-scale power being the key distinction criterion between our $f_\mathrm{esc}$ models, we derive the redshift evolution of the corresponding slope of the 21cm power spectra, $\left(\Delta_\mathrm{21,large} - \Delta_\mathrm{21,small}\right)/\left(k_\mathrm{21,large} - k_\mathrm{21,small}\right)$. This slope reaches the lowest (highest) values for $f_\mathrm{esc}$ models that scale positively (negatively) with stellar mass. These $f_\mathrm{esc}$ model dependent characteristics, particularly the negative slope for $f_\mathrm{esc}$ models scaling positively with stellar mass, provide an avenue to distinguish between different $f_\mathrm{esc}$ dependencies of star-forming galaxies during reionization by means of the evolving 21cm power spectra.

We find that $200$ hour observations with the SKA1-Low allow us to distinguish between different $f_\mathrm{esc}$ models with similar optical depth values ($\tau$). For $\tau\simeq0.055$, measurements of the $21$cm power spectra between redshifts $z\simeq7.5$ and $z \simeq 8.5$ provide the highest constraining power. 
However, taking the uncertainties of the optical depth measurements into account, it is crucial to pinpoint the redshifts of the maximum and minimum $21$cm power spectrum slope. This requires not only $21$cm power spectra measurements for a larger redshift interval ($z\simeq7$ to $z\simeq10$), but also higher accuracy. We find that $800$ hour observations with SKA1-Low will enable us to detect (1) the amplitude and redshift of the maximum slope and (2) most importantly the sign of the slope. A positive slope throughout reionizaton will hint at a scenario where the $f_\mathrm{esc}$ values decrease with the stellar mass of the galaxies, while a negative slope during the overlap phase of reionization will indicate that $f_\mathrm{esc}$ increases with the stellar mass of galaxies. 

In summary, measuring the relation between the large- and small-scale power of the 21cm power spectrum with SKA will allow us to derive constraints on the dependence of the escape fraction of ionizing photons on stellar mass, increasing our knowledge on high-redshift galaxy properties critically. Nevertheless, emission line detections of high-redshift galaxies and possibly higher order statistics such as the 21cm bispectrum may be still required to pin down this property of the sources of reionization.

\section*{Acknowledgements}

We would like to thank Cathryn Trott for providing the baseline distributions and sensitivity of the Square Kilometre Array. We would also like to thank Emma Ryan-Weber for useful discussions. Finally, we thank Simon Mutch for helpful comments during the final stages of preparation.  JS and AH have been supported under the Australian Research Council's Discovery Project funding scheme (project number DP150102987). AH acknowledges additional support from the European Research Council's starting grant ERC StG-717001 ``DELPHI''.  Parts of this research were conducted by the Australian Research Council Centre of Excellence for All Sky Astrophysics in 3 Dimensions (ASTRO 3D), through project number CE170100013.  The Semi-Analytic Galaxy Evolution (\texttt{SAGE}) model used in this work is a publicly available codebase that runs on the dark matter halo trees of a cosmological N-body simulation.  It is available for download at https://github.com/darrencroton/sage.  

\bibliographystyle{mnras}
{\footnotesize
	\setlength{\itemsep}{1pt}
	\begin{spacing}{0.01}
		\bibliography{manuscript_arxiv}
	\end{spacing}	
}

\end{document}